\documentclass[aps,prd,floatfix,notitlepage,twocolumn,superscriptaddress,nofootinbib,amsmath,amssymb]{revtex4-2}

\usepackage{graphicx}
\usepackage{amsmath}
\usepackage{amssymb,amsfonts}
\usepackage{color}
\usepackage[colorlinks,linkcolor=blue,citecolor=blue,urlcolor=blue]{hyperref}
\usepackage{subfigure} 
\def\vk{|{\mathbf k}|}

\begin{document}

\title{DM induced neutron disappearance as the origin of the LZ nuclear recoil event}

\author{P.\,Uttayarat}
\email{patipan@g.swu.ac.th}
\affiliation{Department of Physics, Srinakharinwirot University, 114 Sukhumvit 23rd Rd., Wattana, Bangkok 10110, Thailand}

\author{J.\,Julio}
\email{julio@brin.go.id}
\affiliation{National Research and Innovation Agency, Kawasan Sains dan Teknologi B.\,J.\,Habibie, South Tangerang 15314, Indonesia}

\author{R.\,Primulando}
\email{reinard.primulando@unpar.ac.id}
\affiliation{Center for Theoretical Physics, Department of Physics, Parahyangan Catholic University, Jl. Ciumbuleuit 94, Bandung 40141, Indonesia}


\begin{abstract}
    We investigate dark-matter-induced neutron disappearance as an interpretation of the high-energy nuclear recoil event reported by LUX-ZEPLIN. An incoming dark matter (DM) particle annihilates a
bound neutron into an invisible scalar, leaving a recoiling daughter nucleus. For each accessible daughter state, two-body kinematics fixes the recoil energy in the zero-velocity limit,
producing a line spectrum broadened by halo velocities. We identify parameter regions for benchmark DM masses of $5$ and $50$~GeV that can account for the event near $248$~keV without producing events at lower energy. We examine collider constraints in an illustrative ultraviolet completion and recast Borexino data to constrain accompanying nuclear de-excitation signals. More generally, the DM-induced neutron disappearance provides a framework for interpreting localized nuclear recoil excesses in future direct-detection experiments.
\end{abstract}

\maketitle

\section{Introduction}
The nature of dark matter (DM) remains one of the most outstanding puzzles in modern physics. Among the leading experimental efforts, the LUX-ZEPLIN (LZ) Collaboration operates a dual-phase liquid xenon time projection chamber designed to detect a generic weak scale DM. Recently, based on 2.84 tonne-year exposure, the collaboration reported the observation of a single, highly isolated nuclear recoil (NR) event at an anomalously high energy of approximately $248\pm23 \text{(stat)} \pm 23$ (sys) keV~\cite{LZ:2026axp}. Characterized by a global statistical significance of roughly $2.6\sigma$, this single event stands out cleanly from the expected low-energy background models. 

The complete absence of a corresponding signal at low energies poses a severe challenge to standard elastic DM models, which generically predict rising NR rates toward the detector threshold. To reconcile this feature, initial interpretations have leaned toward inelastic dark matter frameworks~\cite{Fan:2026kxx,Freese:2026sga,Wu:2026nhi,Yin:2026jnn,Nomura:2026qyq,Visinelli:2026kgt,Smirnov:2026aqk,Du:2026guj,Kotlarski:2026pep,Ahmed:2026qjg,Bisal:2026khf,Cheung:2026byg,Langhoff:2026ujr,Chatterjee:2026scv,Yamashita:2026ump,deLima:2026shq,Lee:2026wof,Das:2026uyy,Okada:2026eol,Du:2026lpa,Yuan:2026djt,Zhu:2026dag,Qi:2026vyp,Kumar:2026lgi,Wang:2026ytg,Bandyopadhyay:2026gjw,Borah:2026zwf,Lee:2026xxh,Lee:2026jxl,Okada:2026upm,Asadi:2026iot,He:2026hqz,Su:2026rwz,DiMauro:2026ldr,McCabe:2026crm,Dent:2026bji,Baer:2026fpy,Fan:2026hzw}. In these scenarios, the DM particle must undergo an endothermic transition to a heavier excited state, naturally raising the energy threshold and suppressing low-energy scattering. The recoil spectrum in this scenario typically extends into the side-band region where LZ report zero NR event, posing a challenge inelastic scattering explanation~\cite{Rodd:2026tyn}. Alternative interpretation of LZ NR event invokes momentum suppressed scattering which automatically suppresses the recoil rate at low energy~\cite{Unwin:2026rdp,Khan:2026nwp,Liang:2026coz,Alhazmi:2026efz,Kannike:2026qyl,Heikinheimo:2026kwp,Elahi:2026vlm} without extending the spectrum into the side-band region. These two explanations typically give continuous NR spectrum. 

 In this work, we propose an alternative mechanism: DM-induced neutron disappearance, wherein an incoming DM particle $\chi$ annihilates a bound neutron $n$ inside a target xenon nucleus and produces an invisible particle $\phi$ ($\chi + n \to \phi$). After the neutron has been annihilated, the residual nucleus recoils with momentum and energy fixed by the kinematic of the underlying 2--to--1 process and the neutron removal energy of the residual nucleus state. As a result, the NR spectrum is a set of lines, one for each accessible residual state, instead of a continuous spectrum. The line feature of the NR spectrum has also been noted in Ref.~\cite{Lee:2026zbr} which consider NR recoil from the annihilation of two bound neutrons into a dark particle. This neutron disappearance was also raised in Ref.~\cite{Aghaie:2026vsu} by considering the DM-induced neutron disappearance and neutron invisible decay. We note that typically such an invisible decay involving three body decay at the nucleus level resulting in continuous energy spectrum with the peak at lower energy, which is in tension with the LZ results that see no excess of NR events above background. DM-induced nucleon destruction has been studied in several other frameworks~\cite{Strumia:2021ybk,Ge:2024lzy,Bell:2025uup,Davoudiasl:2010am,Davoudiasl:2011fj,Lou:2026idn}.

\section{Dark matter induced neutron disappearance }
\label{sec:model}

\textit{\textbf{Kinematics}}:---Dark matter induced neutron disappearance  is described by the effective operator
\begin{equation}
\mathcal{L}_{\rm eff}=\bar{n}\,(y_s+y_p\gamma^5)\chi\,\phi+{\rm h.c.},
\label{eq:eft}
\end{equation}
where $\chi$ is a Dirac fermion and $\phi$ is a scalar, both neutral under the
standard model gauge group. 

We are interested in the absorption of $\chi$ on a neutron bound in a nucleus,
$\chi+{}^{A}{\rm Xe}\to\phi+{}^{A-1}{\rm Xe}$, which converts the target into
a definite daughter state rather than scattering it
elastically~\cite{Dror:2019onx,Dror:2019dwc}. We work in the impulse
approximation, in which the struck neutron carries momentum $\mathbf k$ and
removal energy $E_m$ distributed according to the hole spectral function
$S^A_h(\mathbf k,E_m)$~\cite{Benhar:1994hw,Benhar:2006wy}. For a spectator residual
system, energy conservation fixes
\begin{equation}
M_f=M_A-m_n+E_m,\quad E_R=\sqrt{M_f^2+|{\mathbf k}|^2}-M_f,
\label{eq:offshell}
\end{equation}
where $M_f$ is the mass of the daughter including its excitation energy ($E_f^*$) and
$E_R$ is its recoil energy. 
This implies the struck neutron momentum is fixed to the recoils energy,  $|{\mathbf k}| = \sqrt{2M_fE_R}$.
For later convenience, we introduce the bound neutron energy 
\begin{equation}
    \tilde E_n=m_n-E_m-E_R.
\end{equation}

The DM-induced disappearance  cross-section on such a neutron is
\begin{align}
\hat\sigma\,v=\hat\sigma_0\,\delta\!\left(s-m_\phi^2\right),~~\text{with}~~
\hat\sigma_0=\frac{\pi\,\overline{|\mathcal M|^2}}{2E_\chi\tilde E_n},
\label{eq:sigmahat}
\end{align} 
where $v$ is the dark matter velocity in the target frame and
$s=(p_\chi+\tilde p_n)^2$ with $\tilde p_n=(\tilde E_n,\mathbf k)$. 
The reduced cross section $\hat\sigma_0$ carries no mass dimension as the on-shell delta
function has been factored out. The spin averaged matrix element is
\begin{equation}
\overline{|\mathcal M|^2}=\frac{|y_s|^2}{2}\left[s-(m_\chi+\tilde m_n)^2\right]
+\frac{|y_p|^2}{2}\left[s-(m_\chi-\tilde m_n)^2\right],
\label{eq:m2}
\end{equation}
with $\tilde m_n^2=\tilde E_n^2-|{\mathbf k}|^2$. The scalar structure is velocity
suppressed, while the pseudoscalar one survives at vanishing relative velocity. 
\begin{equation}
\hat\sigma_0=\frac{\pi|y|^2}{2}\,
\frac{m_\phi^2-m_\chi^2-\tilde m_n^2}{E_\chi\,\tilde E_n}.
\label{eq:sigma0}
\end{equation}

In the neutron disappearance  process, the energy release is $\Delta=m_\chi+m_n-m_\phi$. Only the nuclear final states $f$ with $E_m<\Delta$ are accessible and their recoil energy is fixed by the 2--to--1 nature of the annihilation process. In the lab frame with the incident DM momentum $\mathbf p_\chi = m_\chi\mathbf v$, the recoil energy can be approximated as
\begin{equation}
  E_R \simeq \frac{m_\phi}{M_f+m_\phi}\,\big(\Delta-E_m\big)
  \;-\;\frac{|{\mathbf p}_\chi|\,|{\mathbf k}|\,\cos\theta}{M_f+m_\phi}.
  \label{eq:ER}
\end{equation}
The above relation shows that $E_R$ is directly related to input parameters ($\Delta$ and $m_\phi$) and a nuclear characteristic $E_m$. 

\begin{figure*}[t]
    \centering
    \subfigure[$m_\chi=5~\mathrm{GeV}$.]{
        \includegraphics[width=0.48\textwidth]
        {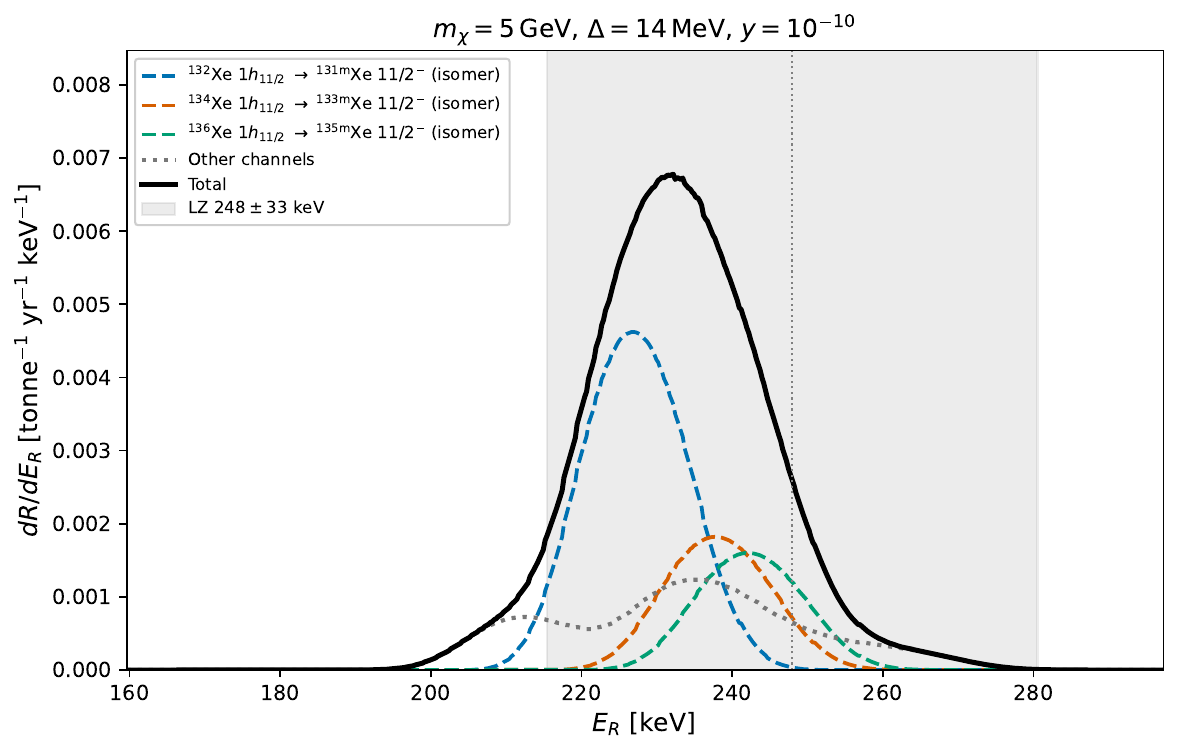}
        \label{fig:bs_light}
    }
    \hfill
    \subfigure[$m_\chi=50~\mathrm{GeV}$.]{
        \includegraphics[width=0.48\textwidth]
        {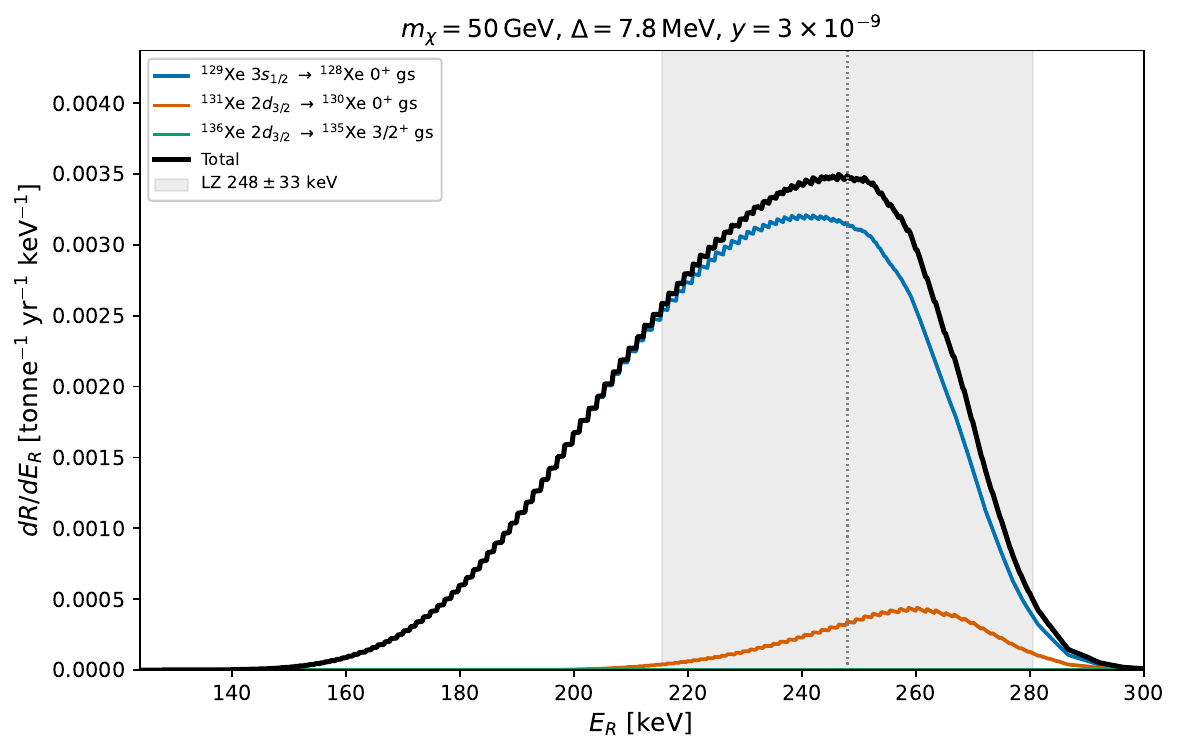}
        \label{fig:bs_heavy}
    }
    \caption{Differential recoil spectra at the two benchmarks:
(a) $\Delta=14$~MeV and (b) $\Delta=7.8$~MeV. Solid curves denote ground-state channels and dashed curves denote various channels. The three largest contributions are shown individually, while the remaining contributions are summed and shown as a dotted curve. The gray band marks the reported LZ recoil
energy, $248\pm32$~keV.}
    \label{fig:benchmark_spectra}
\end{figure*}

The dark matter velocity leads to the spread in the recoil energy. Averaging the second
term of Eq.~(\ref{eq:ER}) over the isotropic $\mathbf k$ of the spectral
function and over the DM velocity distribution in the lab frame broadens $E_R$ into symmetric distributions of
width
\begin{equation}
  \sigma_{E_R}\;=\;\frac{k\,m_\chi}{M_f+m_\phi}
  \sqrt{\frac{\langle v^2\rangle}{3}},
  \label{eq:width}
\end{equation}
with $\langle v^2\rangle^{1/2}\simeq373$~km/s. This is the
only intrinsic width in the DM induced neutron disappearance  scenario. One should note that the width grows linearly with the input DM mass. This broadening effect is responsible for the merging of discrete transition lines 
into the continuous spectrum of Fig.~\ref{fig:benchmark_spectra}.

\textbf{\textit{Rate}}:---The rate per target isotope $A$ follows from folding Eq.~\eqref{eq:sigmahat} with the DM velocity distribution in the lab frame, $\tilde f(v)$, and the hole spectral function
\begin{equation}
    R^A = \frac{\rho_\chi}{m_\chi}\!\int_0^{v_{\rm esc}}\!\!\!\!d^3v\,\tilde f(\mathbf v)
\!\int\frac{d^3k}{(2\pi)^3}\int\!dE_m
\,\hat\sigma v\,S^A_h\!\left(\mathbf k,E_m\right),
\end{equation}
where $\rho_{\chi}\simeq0.4$ GeV/cm$^3$~\cite{Cirelli:2024ssz} is the local DM density and $S_h^A$ incorporates all relevant nuclear physics factors. The total rate is then given by
\begin{equation}
    R = \sum_A\mathcal T_A R^A,
\end{equation}
with $\mathcal T_A$ representing the number of isotope-$A$ nuclei per unit target mass.

To get a qualitative understanding of the rate, consider the limit where DM is at rest. The integral over $\tilde f(\mathbf v)$ is trivial. Furthermore, the 2--to--1 kinematic of the underlying reaction $\chi n\to \phi$ collapses the $\int d^3k$ to $\mathbf k = \mathbf p_0$, the momentum of the recoiling nucleus. Thus the total rate become a set of lines, one for each accessible final state~$f$. However, effects from finite DM velocity and detector response will broaden the lines into overlap spectra, see Fig.~\ref{fig:benchmark_spectra}.

The strength of each transition is determined by the hole spectral function~\cite{Dickhoff:2004xx} 
\begin{equation}
    S_h(\mathbf k,E_m)=\sum_f\big|\langle\Psi^{A-1}_f|a_{\mathbf k}|\Psi^A_0\rangle\big|^2
  \delta\big(E_m-E_m^f\big).
  \label{eq:sh}
\end{equation}
Using the independent particle approximation~\cite{CiofidegliAtti:1995qe}, we can rewrite it as 
\begin{equation}
    S^A_h\!\left({\mathbf k},E_m\right) = 4\pi\sum_f \mathcal S_f|I_{nl}(\vk)|^2\delta(E_m-E_m^f),
\end{equation}
where $f$ runs over possible daughter nucleus final states and $\mathcal S_f$ is the spectroscopic factor. The single neutron removal overlap function is given by~\cite{Ejiri:2018dun}  
\begin{equation}
I_{n\ell}(\vk)=\int_0^\infty\!\!dr\,r^2R_{n\ell}(r)\,j_\ell(\vk r),
\label{eq:overlap}
\end{equation}
and it is normalized to
\begin{equation}
\frac{2}{\pi}\int_0^\infty\!\!d\vk\,\vk^2\left|I_{n\ell}(\vk)\right|^2=1.
\end{equation}
The spectroscopic factor $\mathcal S_{f}$ counts the number of neutrons in the parent nucleus whose removal results in the daughter state $f$, scaled by the quenching factor $q\simeq0.5-0.7$~\cite{Kramer:2001tr,Aumann:2020mnc}. For the odd $A$ parents, only one valence unpaired neutron is removable so $\mathcal S_f = q$, while for the even-$A$ parents, the removal comes from the filled orbital so $\mathcal S_f=(2j+1)v^2_jq$, with $v_j^2$ the neutron occupancy of the parent
ground state. We take $q=0.6$ throughout and determine $v_j^2$ from the sum rule $\sum_j (2j+1)v_j^2 = N_A-50$. 
The radial function $R_{n\ell}$ in Eq.~(\ref{eq:overlap}) is obtained from a mean field method with the Woods--Saxon potential 
\begin{align}
    V(r) = -\frac{V_0}{1+e^{(r-R)/a}},
\end{align}
with $R=r_0A^{1/3}$; $r_0=1.27$~fm and $a=0.67$~fm. For each final state $f$ the parameter
$V_0$ is fixed by requiring that the eigenvalue associated with eigenfunction $rR_{nl}(r)$ is~$-E_m^f$. 

\textbf{\textit{Benchmark}}:---As a concrete example, we consider two benchmark scenarios with $(m_\chi,\Delta, y )=(5~{\rm GeV},14~{\rm MeV}, 10^{-10}$) and $(50~{\rm GeV},7.8~{\rm MeV}, 3\times 10^{-9})$. The possible Xe transitions are listed in Table~\ref{tab:xe_target}. From Eq.~\ref{eq:ER}, one can see that $\Delta$ determines the relevant target nuclei. The recoil spectra for the two benchmarks are shown in Fig.~\ref{fig:benchmark_spectra}. The curves include the nuclear-recoil efficiency of Ref.~\cite{LZ:2026axp} but not the detector energy resolution. Therefore, the width comes from the sum of the individual line spectrum and the velocity broadening given in Eq.~\eqref{eq:width}, with heavier dark matter resulting in a broader spectra.

For $m_\chi = 5~\text{GeV}$ (Fig.~\ref{fig:bs_light}), a higher value of $\Delta$ allows many targets with higher values of $E_m$ to contribute within the energy window of the reported signal. The targets with lower values of $E_m$, i.e., $^{131}$Xe and $^{129}$Xe, have peaks around $345$ keV and $336$ keV, respectively. These lines are outside the science sample and should show up at the high energy sideband.

For the benchmark with lower value of $\Delta = 7.8$ MeV (Fig.~\ref{fig:bs_heavy}), only targets with lower values of $E_m$, particularly $^{131}$Xe and $^{129}$Xe, can generate  recoils. The other xenon isotopes do not contribute as their their removal energies exceed the available energy release for decay. In contrast to the inelastic dark matter explanation for the excess, only few energy recoils are possible in this scenario. For this particular benchmark, no events are expected in either the lower energy part of signal region and higher energy sideband. Therefore, if the LZ excess persists in future xenon experiments, we can distinguish this scenario from the inelastic dark matter scenario.
 
\begin{figure*}[ht!]
    \centering
    \subfigure[$m_\chi=5~\mathrm{GeV}$.]{
        \includegraphics[width=0.48\textwidth]
        {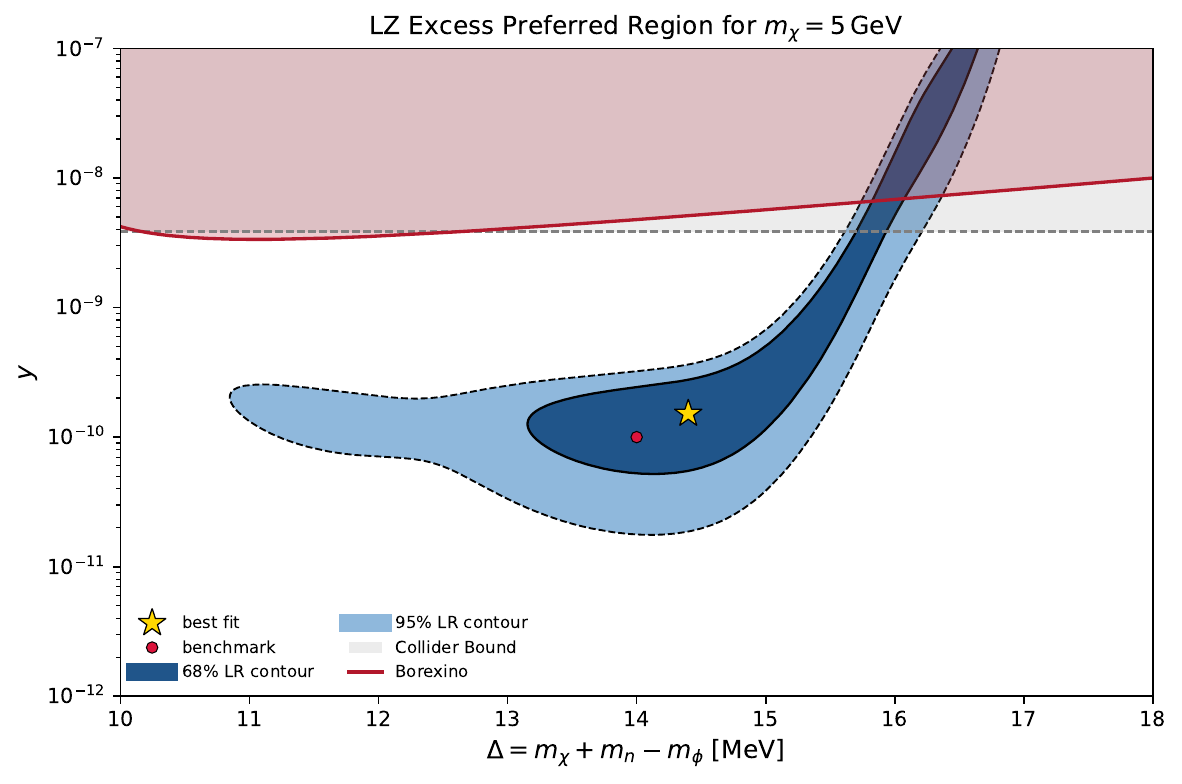}
        \label{fig:benchmark_image_5}
    }
\hfill
    \subfigure[$m_\chi=50~\mathrm{GeV}$.]{
        \includegraphics[width=0.48\textwidth]
        {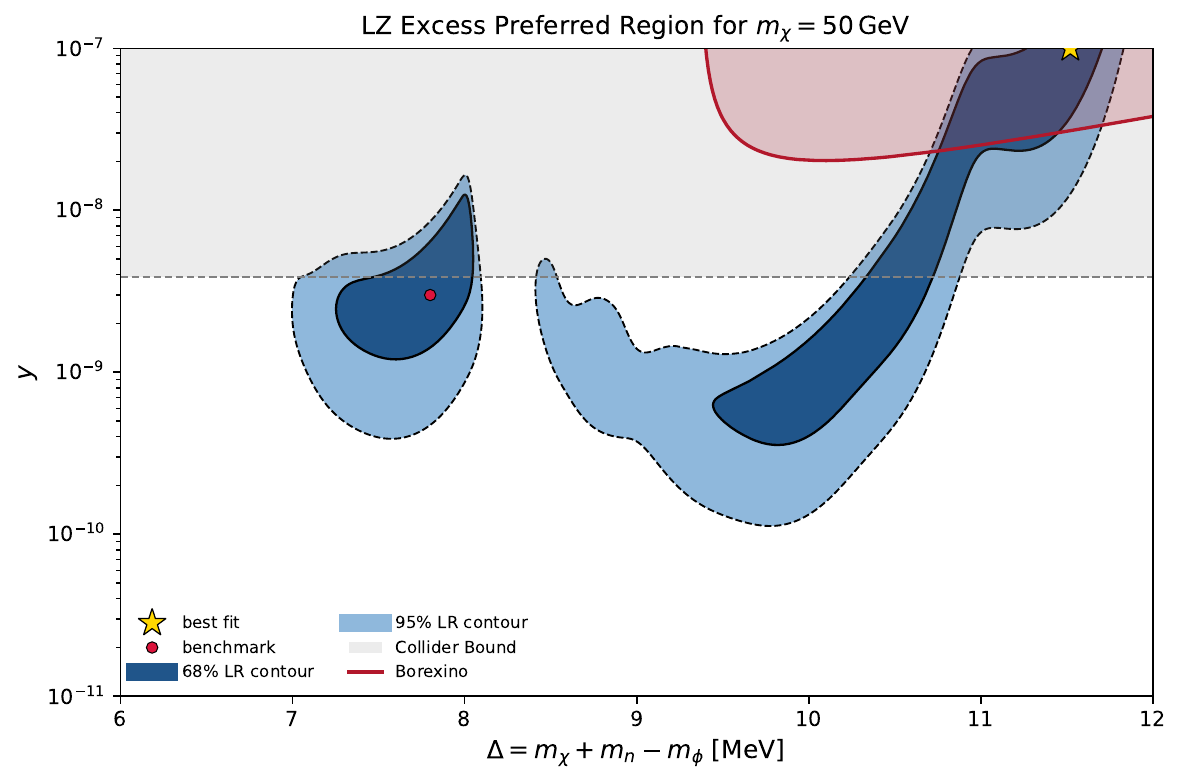}
        \label{fig:benchmark_image_50}
    }
    \caption{The preferred region. The star is the best fit while the circle is the benchmark.}
    \label{fig:benchmark_image}
\end{figure*}

\textbf{\textit{Preferred region}}:---We calculate the preferred region in the $(y,\Delta)$ plane for explaining the LZ excess. For a selected signal region containing a single observed event, we use the extended unbinned likelihood
\begin{equation}
 \mathcal L(y,\Delta)\propto
 e^{-[\mu_s(y,\Delta)+\mu_b]}
 \left[s(E_{\rm LZ};y,\Delta)+b(E_{\rm LZ})\right],
 \label{eq:likelihood}
\end{equation}
where $\mu_s(y,\Delta)$ and $\mu_b$ are the expected numbers of signal and background events in the selected signal region, respectively. The functions $s(E;y,\Delta)$ and $b(E)$ denote the corresponding differential expected event counts per unit recoil energy at the observed energy $E_{\rm LZ}=248~{\rm keV}$. We take flat background distribution along the signal region. Fig.~\ref{fig:benchmark_image} shows the 68\% and 95\% preferred region for both of the benchmark masses, obtained from the likelihood ratio $q = \mathcal{L}/\mathcal{L}_\text{max}$; $\mathcal{L}_\text{max}$ corresponds to the values of $\Delta$ and $y$ that maximize the likelihood, shown as star in each plot. The preferred region in Fig.~\ref{fig:benchmark_image_50} are split into two islands. In the left region, which contains our benchmark point, the signal can only come from $^{131}$Xe and $^{129}$Xe nuclei. In the right island, other nuclei with higher value of $E_m$ can also contribute. 

\section{UV Completion and Bounds}
\label{sec:uvmodel}
\textbf{\textit{UV completion}}:---The effective coupling between $n$, $\chi$ and $\phi$ can be realized in a UV complete model. By introducing a color triplet scalar diquark $\Phi(3,1)_{-1/3}$ and a singlet Dirac fermion $N$, one gets Yukawa interactions
\begin{equation}
\mathcal{L}_{\rm int}\supset y_q\,\overline{u^c_R}d_R\Phi
+y_N\,\Phi^*\bar N d_R+y_\chi\,\bar{N} P_R \chi\phi+{\rm h.c.},
\label{eq:uv}
\end{equation}
where $P_R$ is the right-handed projection operator. Baryon number is conserved by assigning $B_\Phi=-2/3$, $B_N =B_\phi= 1$ and $B_\chi=0$. A similar model has been introduced in Ref.~\cite{Fornal:2018eol} to accommodate the
neutron dark decay $n\to\chi\phi$. At energies below $M_\Phi$, integrating out $\Phi$ introduces a mixing between neutron and the $N$ field, $\mathcal{L}_{\rm eff} \supset (y_q y_N \beta/M_\Phi^2)(\bar{n}N + \bar{N}n)$, where $\beta\equiv\langle0|udd|n\rangle\simeq0.015~{\rm GeV}^3$  is the nucleon to vacuum matrix element obtained from the lattice calculation~\cite{Aoki:2017puj}. This mixing allows a direct coupling between $n$, $\chi$, and $\phi$, i.e., $\mathcal{L}_{\rm eff} \supset 2y\bar{n}P_R\chi\phi + {\rm h.c.}$. By matching onto Eq.~\eqref{eq:eft}, we identify
\begin{align}
    y\equiv y_s = y_p = \frac{y_q y_N y_\chi \beta}{M_\Phi^2(M_N-m_n)}.
\end{align}

The CMS dijet search has placed a lower bound on a scalar diquark mass, with $M_\Phi\ge7.5$ TeV for $y_q$ = 0.3~\cite{CMS:2019gwf}. Meanwhile, the observation of massive neutron stars places $M_N\ge1.2$ GeV~\cite{McKeen:2018xwc}. These constraints give
\begin{equation}
    y \le 2.67\times10^{-10} y_qy_N y_\chi\left(\frac{7.5\,\rm TeV}{M_\Phi}\right)^2\frac{1\, \rm GeV}{(m_N-m_n)}.
\end{equation}
In Fig.~\ref{fig:benchmark_image}, we show the upper bound on $y$ for $y_q =0.3$, $M_\Phi=7.5$ TeV, $M_N=1.2$ GeV, and $y_N=y_\chi=\sqrt{4\pi}$. For this fix UV parameter space, bound is independent of $m_\chi$. Although this model dependend bound exlucde some of the parameter space, portions of the LZ preferred regions are still allowed.

\textbf{\textit{Borexino}}:--- The same interaction that induces neutron disappearance in xenon can also act on neutrons in other nuclei, providing complementary constraints from existing experiments. For $\Delta\gtrsim9.38~{\rm MeV}$, the reaction $\chi+{}^{13}{\rm C}\to\phi+{}^{12}{\rm C}^{*}$ can populate the first $2^+$ state of ${}^{12}{\rm C}$, which promptly decays to the ground state by emitting a $4.44~{\rm MeV}$ photon. The resulting signal falls into the energy range of the Borexino high-energy solar neutrino analysis~\cite{Borexino:2017uhp}. We use 1494 live days of Borexino data with an effective target mass of 227.8 tonne for the HER-I samples. To obtain a conservative upper bound on $y$, we assume that all candidates arise from the signal. The 90\% C.L. bounds are shown in red region in Fig.~\ref{fig:benchmark_image}. Above threshold, the Borexino bounds are competitive with those inferred from the UV model and constrain some LZ preferred region, even excluding the best fit for $m_\chi = 50$ GeV.   

\section{Discussion and Conclusion}

We have investigated the DM-induced neutron disappearance inside the xenon nucleus as the origin of the
high-energy LZ nuclear recoil event. 
DM annihilate a bound neutron to produce an invisible particle which recoils against the residual Xe nucleus. The kinematics of the underlying process, combined with the removal energy of the daughter nuclear state, fixes the set of possible recoil energy of Xe nucleus. The resulting NR spectrum is expected to be a set of discrete lines. However, the velocity spread of the incoming DM broaden the spectral lines into the localized continuous spectrum. The energy release in the annihilation process allows GeV-scale DM to generate recoil energy in the order of a few hundred keV. Our benchmarks illustrate two distinct set of possible nuclear transitions that reproduce a NR spectrum with a peak close to the observed event.

The DM-induced neutron disappearance process is indifferent to mother nucleus. This suggests complimentary probes in other nuclei. In particular, the energy release in the neutron disappearance process responsible for LZ event are enough to excite nuclear transition in $^{2}$H, $^{13}$C, $^{17}$O, etc.  In particular, the NR recoil of $^{13}$C, which is present in liquid scintillator, will be accompanied by a prompt photon with energy 4.44 MeV. This process is constrained by Borexino data, and could be further probed by the JUNO experiment.


Finally, in our scenario, the location of the peak spectrum is determined by the energy released by the DM-neutron annihilation process, the DM mass, and the accessible nuclear final state. This makes it a versatile mechanism for explaining possible future localized NR event in the DM direct detection experiments.

\begin{acknowledgments}
R.P. was supported by Direktorat Penelitian dan Pengabdian kepada Masyarakat, Direktorat Jenderal Riset dan Pengembangan, Kementerian Pendidikan Tinggi, Sains dan Teknologi
Republik Indonesia in the year 2025 with contract number 7939/LL4/PG/2025; III/LPPM/ 2025-06/154-PE and 125/C3/DT.05.00/PL/2025. The work of P.U. was supported in part by Thailand NSRF via PMU-B under grant number B39G690007, and via Srinakharinwirot University under grant number 055/2569.
\end{acknowledgments}

\appendix
\section{Targets for direct detection}
\label{app:xe_channels}

Table~\ref{tab:xe_target} collects every neutron-hole channel in natural
xenon that leaves the daughter either in its ground state or in a
long-lived isomer, so that no prompt de-excitation accompanies the
nuclear recoil. These are the channels that can contribute to the LZ
signal. Removal energies are $E_m=S_n+E_f^*$, with the neutron
separation energies $S_n$ of the parent and the daughter level energies
$E_f^*$ taken from AME2020~\cite{Wang:2021xhn} and
NUBASE2020~\cite{Kondev:2021lzi}; isotopic abundances are the IUPAC
representative values~\cite{Meija:2016xxx}, and are mole fractions, so
the number of isotope-$A$ nuclei per unit target mass is
$\mathcal T_A=x_A N_{\rm A}/\overline{M}$ with
$\overline{M}=131.29$~g/mol the mean molar mass of natural xenon.

\begin{table}[t!]
\caption{Non-$\gamma$-tagged neutron-removal channels in natural
xenon. The daughter half-life is quoted for every channel; ``stable''
marks a stable daughter. Every quoted $T_{1/2}$ refers to the subsequent
weak or isomeric decay of the daughter. Isomeric daughters
are denoted $^{A\rm m}$Xe. $E_m=S_n+E_f^*$ is the removal energy, which
is also the minimum energy release $\Delta$ at which the channel is open,
and $x_A$ is the natural abundance of the parent.}
\label{tab:xe_target}
\begin{ruledtabular}
\footnotesize
\setlength{\tabcolsep}{3.5pt}
\begin{tabular}{lllccc}
parent & orbital & daughter & $T_{1/2}$ & $E_m$ & $x_A$ \\
       &         &          &           & [MeV] & [\%]  \\
\hline
$^{131}$Xe & $2d_{3/2}$  & $^{130}$Xe $0^+$         & stable    &  6.604 & 21.232 \\
$^{129}$Xe & $3s_{1/2}$  & $^{128}$Xe $0^+$         & stable    &  6.907 & 26.401 \\
$^{136}$Xe & $2d_{3/2}$  & $^{135}$Xe $3/2^+$       & 9.14~h    &  8.087 &  8.857 \\
$^{136}$Xe & $1h_{11/2}$ & $^{135\rm m}$Xe $11/2^-$ & 15.29~min &  8.614 &  8.857 \\
$^{134}$Xe & $2d_{3/2}$  & $^{133}$Xe $3/2^+$       & 5.25~d    &  8.554 & 10.436 \\
$^{134}$Xe & $1h_{11/2}$ & $^{133\rm m}$Xe $11/2^-$ & 2.198~d   &  8.787 & 10.436 \\
$^{132}$Xe & $2d_{3/2}$  & $^{131}$Xe $3/2^+$       & stable    &  8.937 & 26.909 \\
$^{132}$Xe & $1h_{11/2}$ & $^{131\rm m}$Xe $11/2^-$ & 11.84~d   &  9.101 & 26.909 \\
$^{130}$Xe & $3s_{1/2}$  & $^{129}$Xe $1/2^+$       & stable    &  9.256 &  4.071 \\
$^{130}$Xe & $1h_{11/2}$ & $^{129\rm m}$Xe $11/2^-$ & 8.88~d    &  9.492 &  4.071 \\
$^{128}$Xe & $3s_{1/2}$  & $^{127}$Xe $1/2^+$       & 36.3~d    &  9.611 &  1.910 \\
$^{126}$Xe & $3s_{1/2}$  & $^{125}$Xe $1/2^+$       & 16.9~h    & 10.018 &  0.089 \\
$^{124}$Xe & $3s_{1/2}$  & $^{123}$Xe $1/2^+$       & 2.08~h    & 10.490 &  0.095 \\
\end{tabular}
\end{ruledtabular}
\end{table}

\bibliographystyle{apsrev4-2}
\bibliography{references}

\end{document}